# Structural stability and electronic properties of $SP^3$ type silicon nanotubes

*Alon Hever, Jonathan Bernstein, and Oded Hod*[*]

Department of Chemical Physics, School of Chemistry, The Sackler Faculty of Exact Sciences, Tel-Aviv University, Tel-Aviv 69978, Israel

## Abstract

A density functional theory study of the structural and electronic properties and relative stability of narrow $SP^3$ silicon nanotubes of different growth orientations is presented. All nanotubes studied and their corresponding wire structures are found to be meta-stable with the wires being more energetically stable. Silicon nanotubes show a dramatic bandgap increase of up to 68% with respect to the corresponding wires. Furthermore, a direct relation between the bandgap of the system and the molar fraction of the passivating hydrogen contents is found. These results suggest that by careful control over their crystallographic growth orientation, dimensions, and chemical composition it should be possible to design and fabricate silicon nanotubes with desired electronic properties.

**Introduction**

Silicon nanowires (SiNWs) and nanotubes (SiNTs) have recently raised as promising candidates for basic components of future nano devices.[1] their low dimensionality leads to quantum confinement effects which can be harnessed to control their electronic properties.[2-7] This opens the way for many possible applications including electronic components such as nanoscale field-effect – transistors,[8-15] high sensitivity chemical and biological detectors,[16-21] and optoelectronic devices.[22-25]

In recent years, several methods have been developed for SiNWs fabrication including laser ablation metal-catalytic vapor-liquid-solid methods,[26-32] oxide-assisted catalyst-free approaches,[33-35] as well as solution based techniques[28]. These methods yield wires with different crystallographic orientations and dimensions scaling down to diameters which are in the single nanometer regime.[26-28,33] The obtained wires often consist of an oxide layer which can be removed and replaced by hydrogen termination,[33] Alternatively, hydrogen passivation can be achieved by using $H_2$ as a carrier gas in the chemical vapor deposition procese.[32] Many efforts have been further invested in the synthesis of SiNTs,[36-39] however only recently techniques enabling the robust synthesis of crystalline SiNTs have emerged. Ben Ishai and Patolsky[40] have reported the formation of robust and hollow single-crystalline silicon nanotubes, with uniform and well-controlled inner diameter, ranging from as small as 1.5 up to 500 nm, and controllable wall thickness and chemical passivation. Quitoriano *et al.*[41] have also reported single-crystalline SiNTs growth using vapor-liquid-solid techniques.

Several theoretical studies have investigated the structural and electronic properties of SiNWs as a function of crystallographic growth direction, radial dimensions, chemical doping, surface passivation, and surface reconstruction.[2-5,42-52] To this end, different computational methods have been utilized to treat the electronic structure of the various systems including semi-empirical methods,[51] density functional based tight binding (DFTB) calculations,[45,50] density functional theory (DFT) calculations,[4-7,48,51-52] as well as many-body perturbation theory within the GW approach.[48] These studies have indicated that for SiNWs with diameter smaller than 4 nm quantum size effects become dominant. As can be expected, decreasing the diameter of the NW was found to results in an increase of the material bandgap accompanied, in some systems, by a transition from an indirect to a direct gap.[3-5,53]

Theoretical studies of SiNTs have recently emerged focusing on single-walled $SP^2$ type silicone analogues of carbon nanotubes (CNTs).[54-56] The electronic structure of these systems was found to be chirality dependent with transition from metallic to semi-conducting depending on the chiral vector orientation similar to the case of their carbon counterparts.[54-55] It was further shown that a slightly distorted structure of single-walled SiNTs where the Si-Si bonds have a somewhat enhanced $SP^3$ character is more stable that the pristine CNT-like structure.[55,57-59] Controlling the structural and electronic properties of SP$^2$ SiNTs via different hydrogenation schemes has also been explored.[59] Beyond the SP$^2$ model, prismane-like[60-61] SiNTs constructed from stacked and covalently bonded square, pentagonal, and hexagonal silicon rings have been investigated predicting a metallic character.[62]

In the present study, inspired by the work of Patolsky *et al.*[40] and Quitoriano *et al.*[41], we present a first-principles study of the structural stability and electronic properties of hydrogen passivated narrow SP$^3$ type SiNTs with a wall thickness of a few atomic layers. We consider a set of SiNWs and SiNTs of different crystallographic orientation including the [100], [110], [111], and the [112] directions (see Fig.1). The SiNWs unit cells have been carved out of bulk silicon and passivated to avoid dangling bonds using hydrogen atoms. To obtain the corresponding SiNTs, the inner core of the SiNWs has been further removed and the resulting new dangling bonds have been passivated using more hydrogen atoms.

**Methods**

All DFT calculations have been performed using the *Gaussian* 09 suite of programs[63]. Three different functional approximations have been considered, namely, the local density approximation (LDA),[64-65] the PBE realization of the generalized gradient approximation,[66-67] and the screened exchange hybrid density functional, HSE.[68-71] The latter functional has been tested for a wide set of materials and was shown to accurately reproduce experimental structural parameters and bandgaps.[72] Initial geometry optimizations have been performed using the LDA with the 3-21G atomic centered Gaussian basis set. Further geometry relaxation has been performed for each functional approximation separately using the double-$\zeta$ polarized 6-31G$^{**}$ basis set.[73] The radial dimensions of the different SiNTs and SiNWs are summarized in Table 1. Coordinates of the fully relaxed structures can be found in the supplementary material.

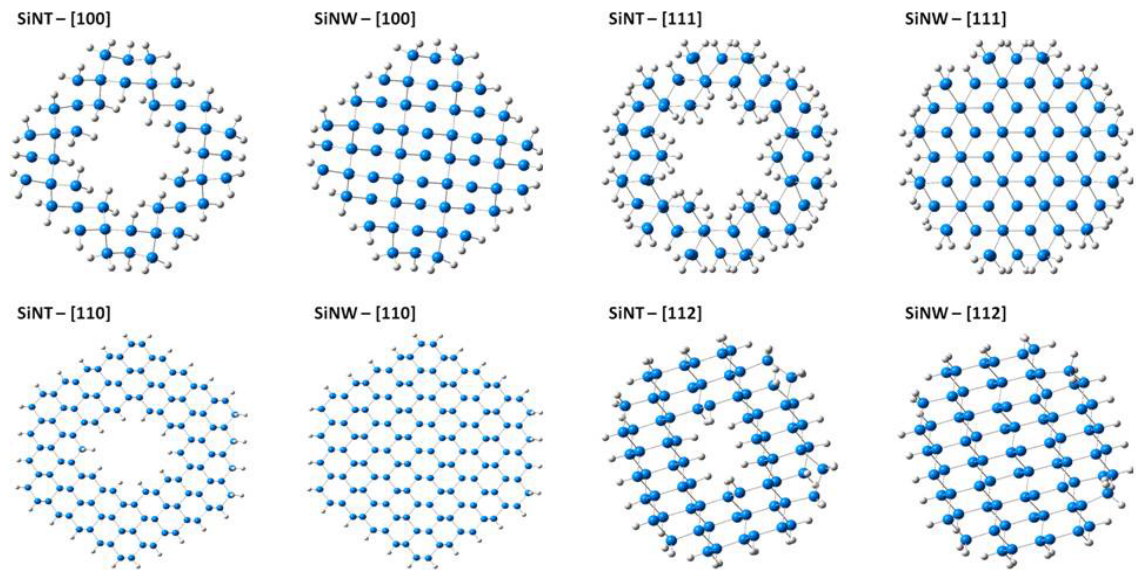

Fig. 1: *Schematic representation of various silicon SiNWs and SiNTs carved out of bulk silicon along the [100],[110],[111] and [112] crystallographic orientations.*

| Growth orientation | Diameter [nm] | | | | | | | | |
|---|---|---|---|---|---|---|---|---|---|
| | SiNW | | | Outer SiNT | | | Inner SiNT | | |
| | LDA | PBE | HSE | LDA | PBE | HSE | LDA | PBE | HSE |
| [100] | 1.50 | 1.50 | 1.51 | 1.50 | 1.54 | 1.53 | 0.90 | 0.94 | 0.93 |
| [110] | 2.64 | 2.67 | 2.65 | 2.63 | 2.66 | 2.65 | 1.37 | 1.39 | 1.38 |
| [111] | 1.54 | 1.56 | 1.55 | 1.55 | 1.57 | 1.56 | 1.34 | 1.38 | 1.37 |
| [112] | 1.55 | 1.56 | 1.56 | 1.57 | 1.59 | 1.58 | 1.15 | 1.16 | 1.16 |

*Table 1: Average radial dimensions of the SiNTs and SiNWs structures optimized using the LDA, PBE, and HSE exchange-correlation functional approximations and the 6-31G\*\* basis set.*

Convergence tests of the electronic structure calculations with respect to the size of the basis set have been performed for the SiNT [100] system. Table 2 presents the bandgap of this system as calculated using the 6-31G\*\* and 6-311G\* basis sets. As can be seen, for all functional approximations considered the bandgaps calculated using the two basis sets are converged within less than 1%.

| | Bandgap [eV] | | |
|---|---|---|---|
| **Functional** | **6-31G\*\*** | **6-311G\*\*** | Δ [%] |
| RSVWN5 | 2.94 | 2.97 | 0.89 |
| RPBEPBE | 3.07 | 3.08 | 0.28 |
| RHSE1PBE | 3.88 | 3.89 | 0.32 |

*Table 2: Bandgap of SiNT [100] as calculated using the LDA, GGA and HSE exchange-correlation functional approximations with the 6-31G\*\* and 6-3111G\* atomic centered Gaussian basis sets.*

**Results and discussion**

We start by analyzing the relative structural stability of the different NWs and NTs shown in Fig. 1. As the SiNWs and SiNTs structures have different chemical compositions the cohesive energy per atom does not provide a suitable measure for the comparison of their relative stability. Therefore, we define the Gibbs free energy of formation ($\delta G$) for SiNT and SiNW as:[5,45,74]

$$\delta G(\chi_{Si}, \chi_H) = E(\chi_{Si}, \chi_H) - \chi_{Si}\mu_{Si} - \chi_H\mu_H \qquad (1)$$

where $E(\chi_{Si}, \chi_H)$ is the cohesive energy per atom of a SiNW/T of a given composition, $\chi_i$ is the molar fraction of atom $i$ ($i$=Si, H) in the system with $\sum_i \chi_i = 1$, and $\mu_i$ is the chemical potential of element $i$. Here, we choose $\mu_H$ as the binding energy per atom of the ground state of the $H_2$ molecule and $\mu_{Si}$ as the cohesive energy per atom of bulk silicon. This definition allows for a direct energy comparison between SiNW/NT with different chemical compositions, where negative values represent stable structures with respect to the constituents. It should be stressed that this treatment gives a qualitative measure of the relative stability while neglecting thermal and substrate effects and zero point energy corrections.[45]

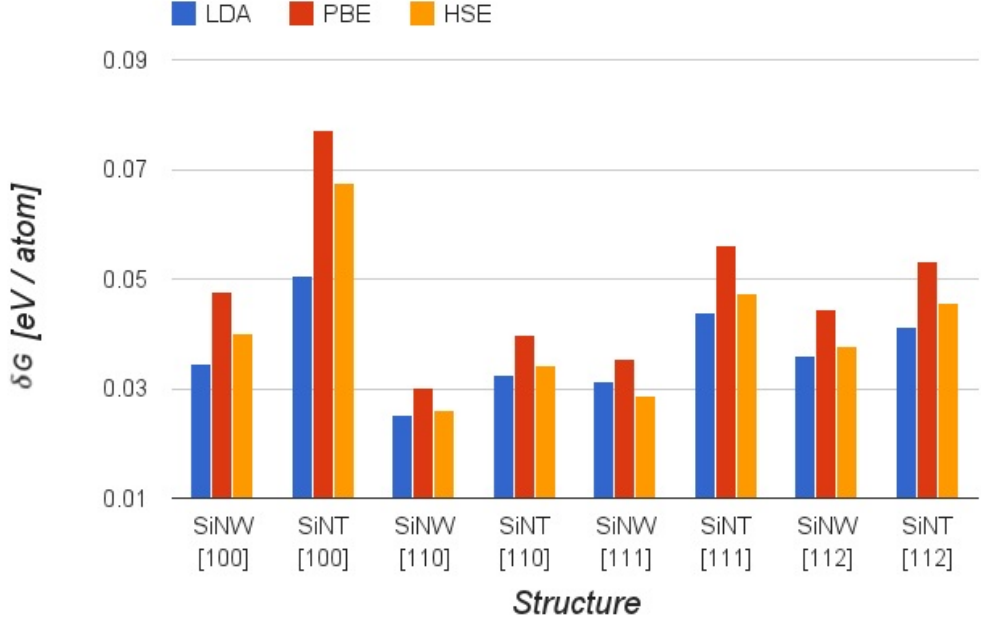

Fig. 2: *δG as calculated using Eq. 1 for the different SiNWs and SiNTs*.

Figure 2 compares the calculated *δG* values for the different SiNWs and SiNTs studied using the LDA, PBE, and HSE exchange-correlation functional approximations. Interestingly, for all functional approximations considered all systems present moderate positive values suggesting that the different structures are meta-stable. This observation is consistent with the results of Aradi *et al.* [45] and Vo *et al.* [5] using a similar method to evaluate the relative stability of other silicon nanowires. The PBE results tend to predict slightly higher *δG* values than the LDA and the HSE functionals. Interestingly, for the [110] NW our calculated *δG* values are considerably smaller than the value calculated by Aradi *et al.*[45] for a narrower system indicating that the stability of the wires should increase with increasing diameter. When comparing the NWs to the NTs we find that for a given growth orientation and radius the SiNWs are more stable than the corresponding SiNTs. We attribute this behavior to the increased surface area of the NTs enhancing surface effects that reduce the stability of the system. In order to support this claim, we plot in Figure 3 the *δG* values of the different systems as a function of the molar fraction of the hydrogen content, $\chi_H$. As the surface is passivated with hydrogen atoms this parameter serves as a measure for the surface to volume ratio. It is clearly evident that as the hydrogen molar fraction reduces and the systems approach bulk silicon (both in terms of chemical composition and in terms of surface effects) *δG* decreases thus indicating on an increased stability.

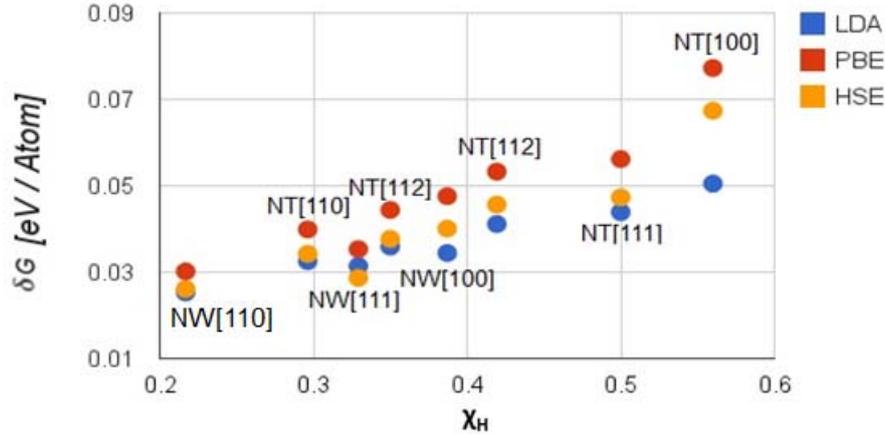

Fig. 3: *δG as calculated from Eq. 1 vs. the hydrogen molar fraction $\chi_H$.*

Apart from the surface to volume ratio and overall chemical composition, other factors such as surface reconstruction and type of passivation as well as steric considerations may influence the relative stability of the different structures. The stability analysis presented in Fig. 2 suggests that among all NWs and NTs studied the [110] growth orientation is the most stable followed by the [111] structures. This is in contrast to the findings of Vo *et al*.[5] suggesting that for temperatures lower than 822K the [111] NW is the most stable structure followed by the [110] and the [100] directions. These differences may result from two factors: (i) The calculations of Vo *et al*. included the zero point energy. As the differences between the *δG* values that we obtained for the [111] and [110] NWs are very small (0.009 eV/atom for the LDA functional used in the study of Vo et al.) zero point energy may change the calculated stability order; (ii) As the current study focuses on the differences between the NWs and NTs, different system diameters were chosen for structures of different growth orientations. Therefore, the effect of system diameter on the relative stability is not taken into account and thus the direct comparison between the relative stabilities of the different NWs and NTs studied herein is limited to the studied structures alone and is not of general nature.

We now turn to discuss the electronic structure of the different systems considered. First, in order to check the validity of our NWs and NTs atomistic models we compare the calculated bandgaps to previously reported results.[3-7,45,49] Fig. 4 presents NW's bandgaps obtained using the LDA functional here and in previous studies as a function of system diameter and crystallographic growth direction. As can be seen, our calculated bandgaps compare well to previously reported results. Similar agreement was obtained for the GGA and HSE results (see supplementary material) indicating the reliability of our calculations. Furthermore, our HSE results for the bandgap of the [110] SiNW of diameter 2.65nm (1.57eV) and the [112] SiNW of diameter 1.56nm (2.11eV) compare well with the experimental measurements of Ma *et al*. for a [110] SiNW of diameter of 3nm (1.5 eV) and a [112] SiNW of diameter of 2 nm (2.28eV), respectively.[33]

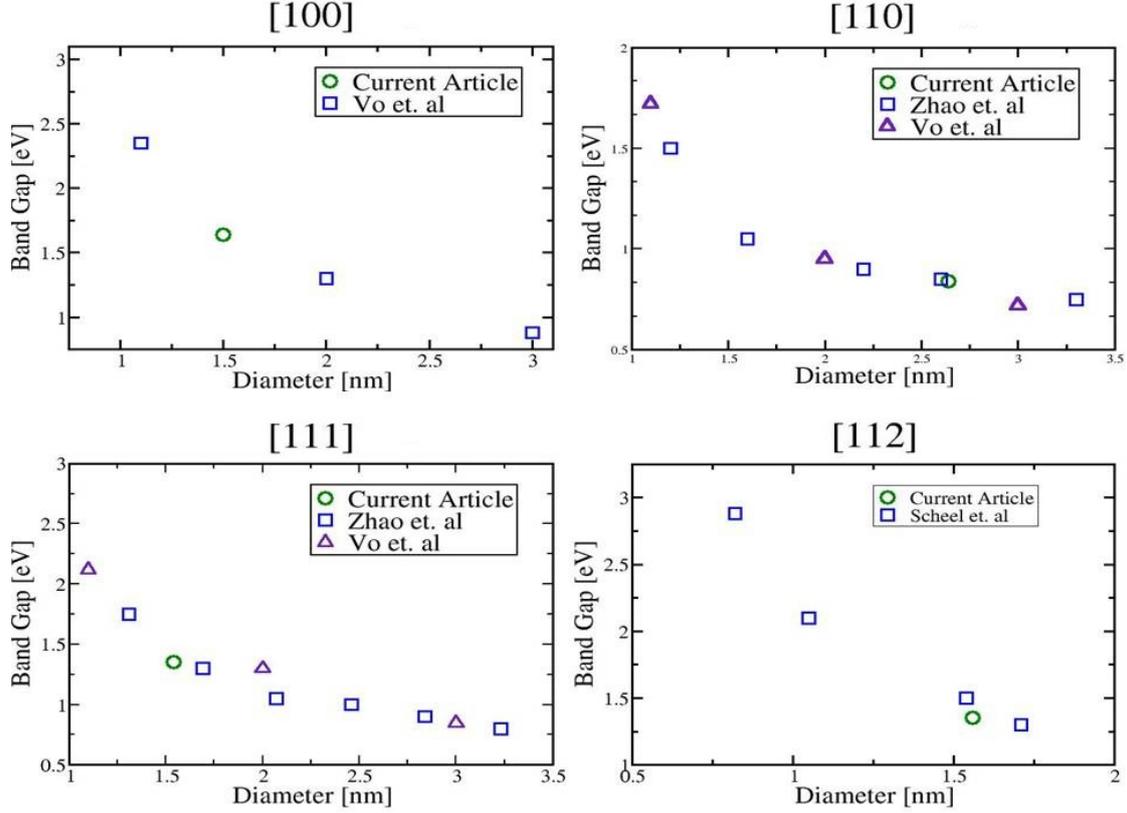

Fig. 4: *Comparison between LDA bandgaps obtained for SiNWs of different growth orientations and reference values.*

Having established the validity of our atomistic models and computational methodology we now focus on the effect of the inner cavity on the electronic properties of SiNTs. Fig. 5 compares the bandgaps of the different SiNWs considered to those of the corresponding SiNTs. As can be clearly seen, for all the different growth orientations studied a dramatic increase of the bandgap is evident when going from the NW to the NT configuration. This is true for all functional approximations utilized. Specifically, for the HSE functional, which is expected to produce the most reliable bandgap values, an increase of 17% , 26%, 61%  and 68% was obtained for the [110], [112], [100] and [111] crystallographic orientation, respectively.

To better understand surface and chemical composition effects on the electronic structure of SiNWs and SiNTs we plot the bandgaps of the different systems considered as a function of the hydrogen molar fraction. This parameter encompasses a complex combination of several chemical and physical factors dictating the overall electronic structure which include surface reconstruction, surface states, chemical composition, surface to volume ratio, and system dimensions. Interestingly, despite this intricate balance of different contributions a direct relation between the bandgap and the hydrogen content is clearly observed. At the limit of zero hydrogen content the HSE bulk silicon bandgap of 1.22 eV is recovered. As the hydrogen content is increased the bandgap increases linearly up to a value of ~4 eV. These results suggest that by careful control over their crystallographic growth orientation, dimensions, and chemical composition it should be possible to fabricate SiNTs with predesigned desired electronic properties.

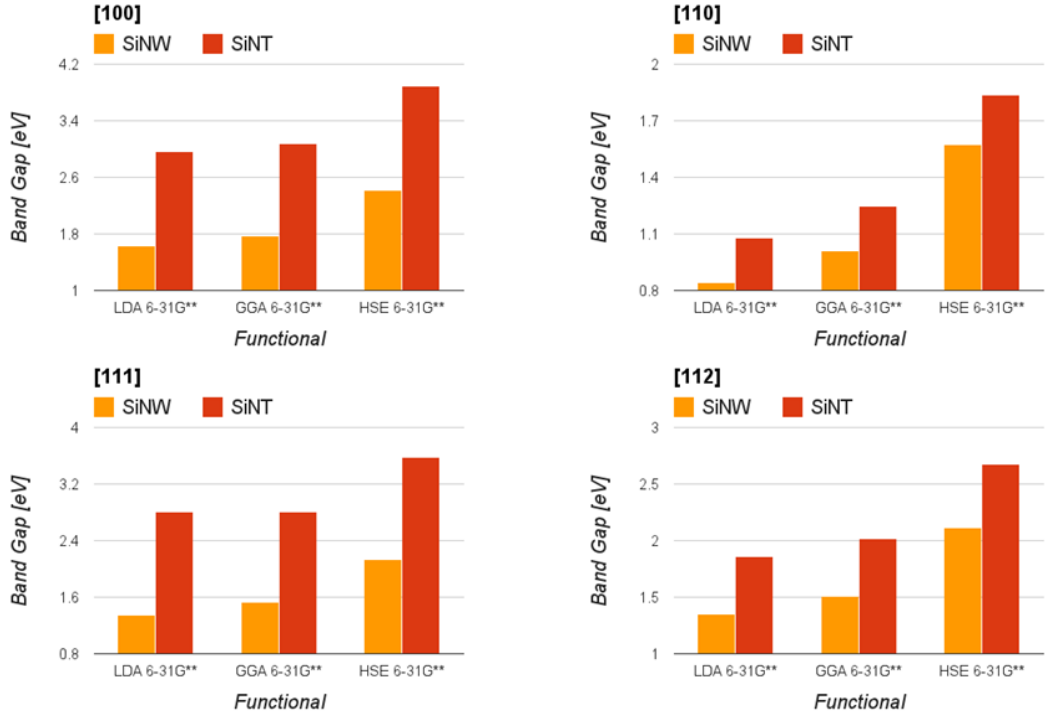

Fig. 5: *Bandgaps of SiNWs of different growth orientations and their corresponding SiNTs as calculated using the LDA, GGA and HSE exchange-correlation functional approximations.*

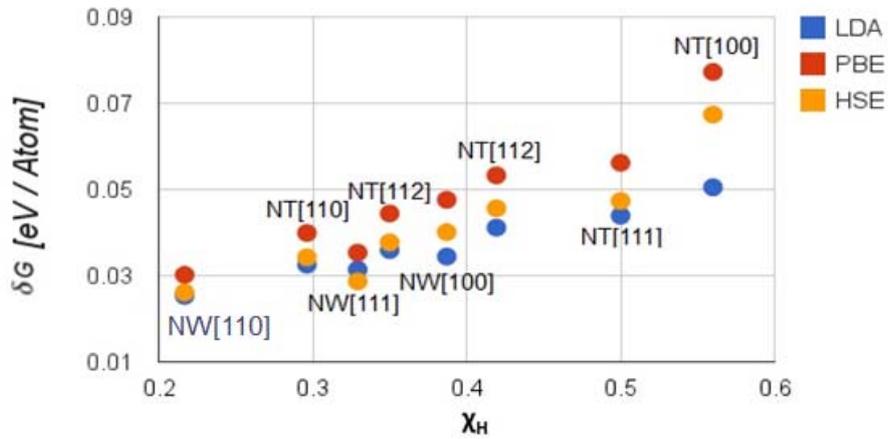

Fig. 6: *Bandgap of SiNWs and SiNTs as a function of the hydrogen molar fraction.*

Finally, in order to obtain a more complete picture of these effects the full band-structure of the different systems has been calculated. In Fig. 7 a comparative view of the band-structures of the different NWs and their corresponding NTs is presented. As can be seen, the overall electronic structure suffers relatively minor modifications when moving from the wire structures to the hollow structures. The main effect of the inner cavity is to lower the valence band maximum and (especially in the case of the [100] system) raise the conduction band minimum thus increasing the overall bandgap. According to our HSE calculations the [100], [110] and [111] NWs are direct bandgap semiconductors while the [112] NW has an indirect gap. Interestingly, all but the [111] NTs have the same type of bandgap as the corresponding NWs. The [111] NT moves from a direct to an indirect gap. This further suggests that not only the value of the bandgap but also its character may be controlled by careful tailoring of the detailed chemical composition and overall structure of the system.

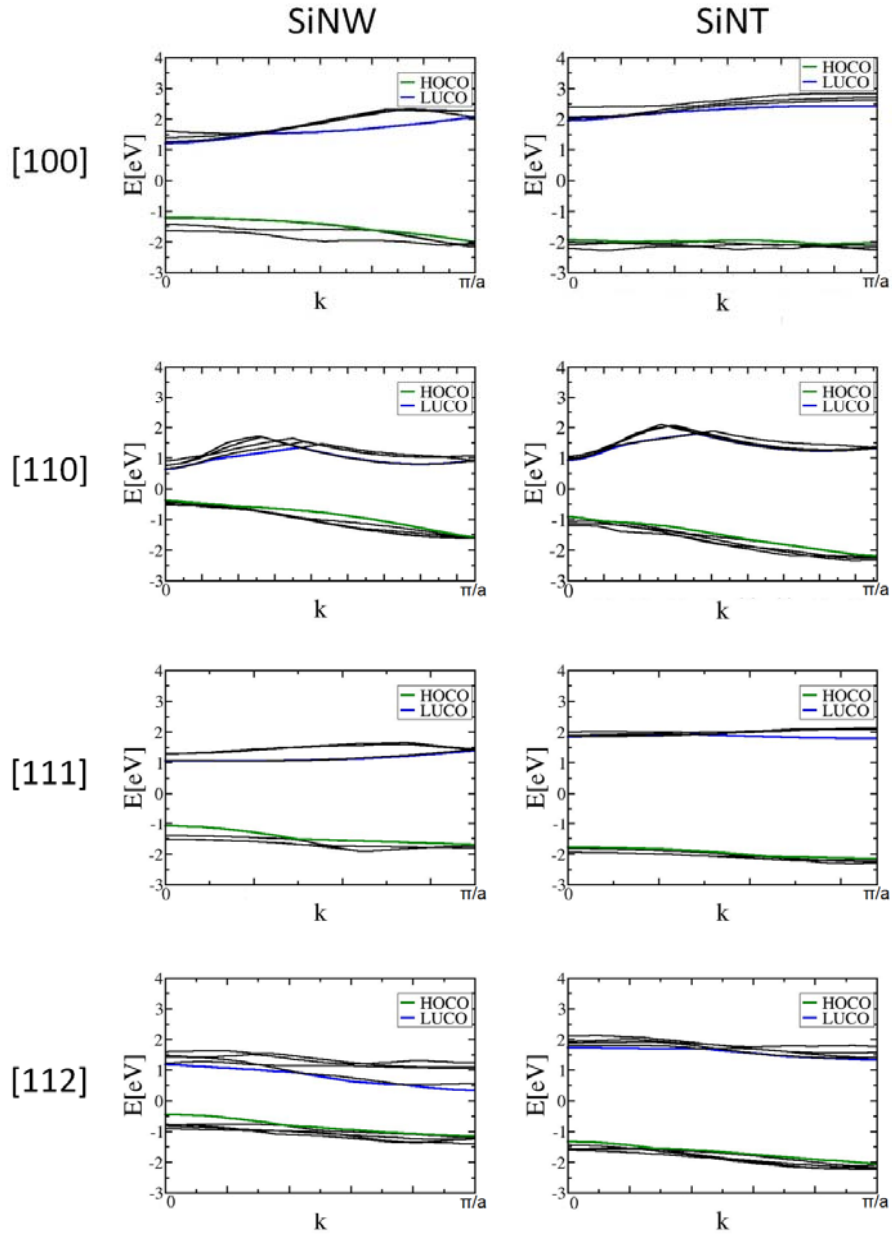

Fig. 7: *Band structures of SiNWs with growth orientations of [100], [110], [111] and [112] and their corresponding SiNTs as calculated at the HSE/6-31G$^{**}$ level of theory. The mid-gap value of all systems is set to zero.*

**Summary**

In this paper we presented a theoretical study of the structural and electronic properties of hydrogen passivated narrow SP$^3$ type SiNTs bearing a wall thickness of a few atomic layers. Four SiNT models with growth orientations along the [100], [110], [111], [112] bulk silicon crystallographic directions were considered. Their energetic stability and electronic properties were compared to the corresponding SiNWs. All SiNT and SiNW considered were found to be meta-stable structures. Furthermore, for all growth orientations studied the SiNTs were found to be less stable than the corresponding SiNWs. When comparing the bandgap of SiNWs and SiNTs of the same growth orientation, the formation of an inner cavity in the wires was found to be accompanied by a significant bandgap increase in the resulting nanotubes. The overall increase in bandgap was found to be 17%, 26%, 61% and 68% for the [110], [112], [100] and [111] crystallographic orientation, respectively. We have found a direct relation between the hydrogen molar fraction and both the structural stability and the bandgap of the different systems. Generally speaking, as the hydrogen contents decreases the structural stability increases and the bandgap decreases indicating that the SiNWs and SiNTs approach the bulk limit. The [100], [110] and [111] NWs were found to be direct bandgap semiconductors while the [112] NW bandgap was of indirect character. Interestingly, all but the [111] NT have kept the bandgap type of the corresponding NWs. Our results suggest that by careful control over their crystallographic growth orientation, dimensions, and chemical composition it should be possible to fabricate SiNTs with predesigned desired electronic properties.

**Acknowledgments**

The authors would like to thank Prof. Fernando Patolsky for helpful discussions. This work was supported by the Israel Science Foundation under grant No. 1313/08, the Center for Nanoscience and Nanotechnology at Tel Aviv University, and the Lise Meitner-Minerva Center for Computational Quantum Chemistry.

# References

(1) Cui, Y.; Lieber, C. M. *Science* **2001**, *291*, 851-853.
(2) Delley, B.; Steigmeier, E. F. *Applied Physics Letters* **1995**, *67*, 2370.
(3) Zhao, Y.; Yakobson, B. I. *Physical Review Letters* **2003**, *91*, 035501.
(4) Leu, P. W.; Shan, B.; Cho, K. *Physical Review B* **2006**, *73*, 195320.
(5) Vo, T.; Williamson, A. J.; Galli, G. *Physical Review B* **2006**, *74*, 045116.
(6) Singh, A. K.; Kumar, V.; Note, R.; Kawazoe, Y. *Nano Letters* **2006**, *6*, 920-925.
(7) Rurali, R.; Aradi, B.; Frauenheim, T.; Gali, A. *Physical Review B* **2007**, *76*, 113303.
(8) Cui, Y.; Zhong, Z.; Wang, D.; Wang, W. U.; Lieber, C. M. *Nano Letters* **2003**, *3*, 149-152.
(9) Duan, X.; Huang, Y.; Cui, Y.; Wang, J.; Lieber, C. M. *Nature* **2001**, *409*, 66-69.
(10) Huang, Y.; Duan, X.; Cui, Y.; Lauhon, L. J.; Kim, K.-H.; Lieber, C. M. *Science* **2001**, *294*, 1313-1317.
(11) Tans, S. J.; Verschueren, A. R. M.; Dekker, C. *Nature* **1998**, *393*, 49-52.
(12) Martel, R.; Schmidt, T.; Shea, H. R.; Hertel, T.; Avouris, P. *Applied Physics Letters* **1998**, *73*, 2447-2449.
(13) Wind, S. J.; Appenzeller, J.; Martel, R.; Derycke, V.; Avouris, P. *Applied Physics Letters* **2002**, *80*, 3817-3819.
(14) Derycke, V.; Martel, R.; Appenzeller, J.; Avouris, P. *Nano Letters* **2001**, *1*, 453-456.
(15) Martel, R.; Derycke, V.; Lavoie, C.; Appenzeller, J.; Chan, K. K.; Tersoff, J.; Avouris, P. *Physical Review Letters* **2001**, *87*, 256805.
(16) Cui, Y.; Q. Wei; H. Park; Lieber, C. M. *Science* **2001**, *293*, 17.
(17) Hahm, J.-i.; Lieber, C. M. *Nano Letters* **2003**, *4*, 51-54.
(18) Gao, Z.; Agarwal, A.; Trigg, A. D.; Singh, N.; Fang, C.; Tung, C.-H.; Fan, Y.; Buddharaju, K. D.; Kong, J. *Analytical Chemistry* **2007**, *79*, 3291-3297.
(19) Patolsky, F.; Zheng, G.; Lieber, C. M. *Nat. Protocols* **2006**, *1*, 1711-1724.
(20) Cheng, M. M.-C.; Cuda, G.; Bunimovich, Y. L.; Gaspari, M.; Heath, J. R.; Hill, H. D.; Mirkin, C. A.; Nijdam, A. J.; Terracciano, R.; Thundat, T.; Ferrari, M. *Current Opinion in Chemical Biology* **2006**, *10*, 11-19.
(21) Kuang, Q.; Lao, C.; Wang, Z. L.; Xie, Z.; Zheng, L. *Journal of the American Chemical Society* **2007**, *129*, 6070-6071.
(22) Li, Y.; Qian, F.; Xiang, J.; Lieber, C. M. *Materials Today* **2006**, *9*, 18-27.
(23) Star, A.; Lu, Y.; Bradley, K.; Grüner, G. *Nano Letters* **2004**, *4*, 1587-1591.
(24) Zhong, Z.; Qian, F.; Wang, D.; Lieber, C. M. *Nano Letters* **2003**, *3*, 343-346.
(25) Wong, H. *Microelectronics Reliability* **2002**, *42*, 317-326.
(26) Zhang, Y. F.; Tang, Y. H.; Wang, N.; Yu, D. P.; Lee, C. S.; Bello, I.; Lee, S. T. *Applied Physics Letters* **1998**, *72*, 1835-1837.
(27) Cui, Y.; Lauhon, L. J.; Gudiksen, M. S.; Wang, J.; Lieber, C. M. *Applied Physics Letters* **2001**, *78*, 2214-2216.
(28) Holmes, J. D.; Johnston, K. P.; Doty, R. C.; Korgel, B. A. *Science* **2000**, *287*, 1471-1473.
(29) Morales, A. M.; Lieber, C. M. *Science* **1998**, *279*, 208-211.
(30) Irrera, A.; Pecora, E. F.; Priolo, F. *Nanotechnology* **2009**, *20*, 135601.
(31) Ross, F. M.; Tersoff, J.; Reuter, M. C. *Physical Review Letters* **2005**, *95*, 146104.
(32) Wu, Y.; Cui, Y.; Huynh, L.; Barrelet, C. J.; Bell, D. C.; Lieber, C. M. *Nano Letters* **2004**, *4*, 433-436.
(33) Ma, D. D. D.; Lee, C. S.; Au, F. C. K.; Tong, S. Y.; Lee, S. T. *Science* **2003**, *299*, 1874.
(34) Wang, N.; Tang, Y. H.; Zhang, Y. F.; Lee, C. S.; Bello, I.; Lee, S. T. *Chemical Physics Letters* **1999**, *299*, 237-242.
(35) Golea, J. L.; Stout, J. D.; Rauch, W. L.; Wangb, Z. L. *Applied Physics Letters* **2000**, *76*, 2346.
(36) Schmidt, O. G.; Eberl, K. *Nature* **2001**, *410*, 168-168.

(37) Jeong, S. Y.; Kim, J. Y.; Yang, H. D.; Yoon, B. N.; Choi, S. H.; Kang, H. K.; Yang, C. W.; Lee, Y. H. *Advanced Materials* **2003**, *15*, 1172-1176.
(38) Chen, Y. W.; Tang, Y. H.; Pei, L. Z.; Guo, C. *Advanced Materials* **2005**, *17*, 564-567.
(39) Crescenzi, M. D.; P. Castrucci; Scarselli, M.; Diociaiuti, M.; Chaudhari, P. S.; Balasubramanian, C.; Bhave, T. M.; Bhoraskar, S. V. *Applied Physics Letters* **2005**, *86*, 231901.
(40) Ben-Ishai, M.; Patolsky, F. *J. Am. Chem. Soc.* **2009**, *131*, 3679-3689.
(41) Quitoriano, N. J.; Belov, M.; Evoy, S.; Kamins, T. I. *Nano Letters* **2009**, *9*, 1511-1516.
(42) Read, A. J.; Needs, R. J.; Nash, K. J.; Canham, L. T.; Calcott, P. D. J.; Qteish, A. *Physical Review Letters* **1992**, *69*, 1232-1235.
(43) Ohno, T.; Shiraishi, K.; Ogawa, T. *Physical Review Letters* **1992**, *69*, 2400-2403.
(44) Hybertsen, M. S.; Needels, M. *Physical Review B* **1993**, *48*, 4608-4611.
(45) Aradi, B.; Ramos, L. E.; Deák, P.; Köhler, T.; Bechstedt, F.; Zhang, R. Q.; Frauenheim, T. *Phys. Rev. B* **2007**, *76*, 35305.
(46) Singh, A. K.; Kumar, V.; Note, R.; Kawazoe, Y. *Nano Letters* **2005**, *5*, 2302-2305.
(47) Rurali, R.; Lorente, N. *Physical Review Letters* **2005**, *94*, 026805.
(48) Zhao, X.; M.Wei, C.; Yang, L.; Chou, M. Y. *Physical Review Letters* **2004**, *92*, 23.
(49) Scheel, H.; Reich, S.; Thomsen, C. *Physica Status Solidi (b)* **2005**, *242*, 2474-2479.
(50) Zhang, R. Q.; Lifshitz, Y.; Ma, D. D. D.; Zhao, Y. L.; Frauenheim, T.; Lee, S. T.; Tong, S. Y. *Journal of Chemical Physics* **2005**, *123*, 144703.
(51) Ng, M.-F.; Zhou, L.; Yang, S.-W.; Sim, L. Y.; Tan, V. B. C.; Wu, P. *Physical Review B* **2007**, *76*, 155435.
(52) Nolan, M.; O'Callaghan, S.; Fagas, G.; Greer, J. C.; Frauenheim, T. *Nano Letters* **2006**, *7*, 34-38.
(53) Delerue, C.; Allan, G.; Lannoo, M. *Physical Review B* **1993**.
(54) Fagan, B.; Baierle, R. J.; R. Mota, A.; Silva, J. R. d.; Fazzio, A. *Physical Review B* **1999**, 15.
(55) Yang, X.; Ni, J. *Phys. Rev. B* **2005**, *72*, 195426.
(56) Barnard, A. S.; Russo, S. P. *J. Phys. Chem.* **2003**, *107*, 7577-7581.
(57) Zhang, R. Q.; Lee, H.-L.; Li, W.-K.; Teo, B. K. *J. Phys. Chem.* **2005**, *109*, 8605-8612.
(58) Zhang, M.; Kan, Y. H.; Zang, Q. J.; Su, Z. M.; Wang, R. S. *Chemical Physics Letters* **2003**, *379*, 81-86.
(59) Seifert, G.; Kohler, T.; Urbassek, H. M.; Hernandez, E.; Frauenheim, T. *Physical Review B,* **2001**, *63*, 193409.
(60) Pour, N.; Altus, E.; Basch, H.; Hoz, S. *Journal of Physical Chemistry C* **2010**, *114*, 10386-10389.
(61) Pour, N.; Altus, E.; Basch, H.; Hoz, S. *Journal of Physical Chemistry C* **2009**, *113*, 3467-3470.
(62) Bai, J.; Zeng, X. C.; Tanaka, H.; Zeng, J. Y. *Proc. Nat. Acad. Sci.* **2004**, *101*, 2665.
(63) Frisch, M. J.; Trucks, G. W.; Schlegel, H. B.; Scuseria, G. E.; Robb, M. A.; Cheeseman, J. R.; Jr., J. A. M.; Vreven, T.; Kudin, K. N.; et al. *Gaussian Inc, Pittsburgh, PA* **2008**.
(64) J. Slater, Quantum Theory of Molecular and Solids, The Self-Consistent Field for Molecular and Solids, Vol. 4, McGraw-Hill, NewYork, 1974.
(65) Vosko, S. H.; Wilk, L.; Nusair, M.; Can *J.Phys* **1980**, *58*, 1200.
(66) Perdew, J. P.; Burke, K.; Ernzerhof, M. *Physical Review Letters* **1998**, *80*, 891-891.
(67) Zhang, Y. K.; Yang, W. T. *Physical Review Letters* **1998**, *80*, 890-890.
(68) Heyd, J.; Scuseria, G. E. *Journal of Chemical Physics* **2004**, *120*, 7274-7280.
(69) Heyd, J.; Scuseria, G. E. *Journal of Chemical Physics* **2004**, *121*, 1187-1192.
(70) Heyd, J.; Scuseria, G. E.; Ernzerhof, M. *Journal of Chemical Physics* **2003**, *118*, 8207-8215.
(71) Heyd, J.; Scuseria, G. E.; Ernzerhof, M.; Phys., J. C. *J. Comp. Phys.* **2006**, *124*, 219924(E).
(72) Heyd, J.; Peralta, J. E.; Scuseria, G. E.; Martin, R. L. *Journal of Chemical Physics* **2005**, *123*, 174101.

(73) Hariharan, P. C.; Pople, J. A. T. *Chim. Acta* **1973**, *213*, 28.
(74) Hod, O.; Barone, V.; Peralta, J. E.; Scuseria, G. E. *Nano Letters* **2007**, *7*, 2295-2299.